\documentclass[10pt]{article}
\usepackage{graphicx}
\usepackage{bm,epsfig,psfrag,color,amsmath}

\newcommand{\ff}{f\hspace{-0.4em}f}
\def\Title#1{\begin{center} {\Large #1 } \end{center}}
\def\Author#1{\begin{center}{ \sc #1} \end{center}}
\def\Address#1{\begin{center}{ \it #1} \end{center}}

\newcommand\pubblock{\rightline{\begin{tabular}{l} Proceedings of the Second Annual LHCP\\ \pubnumber\\
         \pubdate  \end{tabular}}}

\newenvironment{Abstract}{\begin{quotation} \begin{center} 
             \large ABSTRACT \end{center}\bigskip 
      \begin{center}\begin{large}}{\end{large}\end{center} \end{quotation}}

\newenvironment{Presented}{\begin{quotation} \begin{center} 
             PRESENTED AT\end{center}\bigskip 
      \begin{center}\begin{large}}{\end{large}\end{center} \end{quotation}}

\def\Acknowledgements{\bigskip  \bigskip \begin{center} \begin{large}
             \bf ACKNOWLEDGEMENTS \end{large}\end{center}}

\def\beq{\begin{equation}}
\def\eeq#1{\label{#1}\end{equation}}
\def\eeqn{\end{equation}}

\def\beqa{\begin{eqnarray}}
\def\eeqa#1{\label{#1}\end{eqnarray}}
\def\eeqan{\end{eqnarray}}

\let\bar=\overbar

\def\Dslash{\not{\hbox{\kern-4pt $D$}}}
\def\dslash{\not{\hbox{\kern-2pt $\del$}}}

\def\msb{{\bar{\ssstyle M \kern -1pt S}}}

\usepackage{color}

 \newcommand\pubnumber{ }

\newcommand\pubdate{\today}

\def\affiliationa{
Physics Department, \\
New York City College of Technology, The City University of New York,
NY 11201, U.S.A\\
The Graduate School and University Center,\\
 The City University of New York, New York, NY 10016, U.S.A }

  \def\affiliationc{Institute for Particle Physics Phenomenology,\\
   University of Durham DH1 3LE Durham, UK }

\def\affiliationd{School of Physics and State Key Laboratory of Nuclear Physics and Technology,\\
Peking University, Beijing 100871, China}

\begin{document}

\large
\begin{titlepage}
\pubblock

\vfill
\Title{  Boosted Top Quark Pair Production in Soft Collinear Effective Theory  }
\vfill

\Author{ Andrea Ferroglia  }
\Address{\affiliationa}
\Author{ Simone Marzani  }
\Address{\affiliationc}
\Author{ Ben D. Pecjak  }
\Address{\affiliationc}
\Author{ Li Lin Yang }
\Address{\affiliationd}
\vfill
\begin{Abstract}
We review a Soft Collinear Effective Theory  approach to the study of factorization
and resummation of QCD effects in top-quark pair production. In
particular, we consider differential cross sections such as the
top-quark pair invariant mass distribution and the top-quark
transverse momentum and rapidity distributions. Furthermore, we focus
our attention on the large invariant mass and large transverse momentum kinematic
regions, characteristic of boosted top quarks. We discuss the 
factorization of the differential cross section in the double soft gluon emission and small top-quark mass limit, both in Pair Invariant Mass (PIM) and One Particle Inclusive (1PI) kinematics.
The factorization formulas can be employed  in
order to implement the simultaneous resummation of soft emission and small mass effects up to next-to-next-to-leading logarithmic accuracy.
The results are also used to construct improved next-to-next-to-leading order
approximations for the differential cross sections.
\end{Abstract}
\vfill

\begin{Presented}
The Second Annual Conference\\
 on Large Hadron Collider Physics \\
Columbia University, New York, U.S.A \\ 
June 2-7, 2014
\end{Presented}
\vfill
\end{titlepage}
\def\thefootnote{\fnsymbol{footnote}}
\setcounter{footnote}{0}
%

\normalsize 


\section{Introduction}

Scattering processes at hadron colliders typically involve  a hierarchy of scales, often exhibit complicated kinematics, and present soft and collinear singularities which lead to Sudakov double logarithms. Consequently, 
precise theoretical predictions for several observables can be
obtained only after resummation of large logarithmic corrections. For these reasons Soft-Collinear Effective Theory (SCET)  is an ideal tool for the study of these processes.
The general idea is to separate the various scales that are present in the process which one wants to study. 
In order to achieve this goal,  one works in  kinematic regions of the phase space known to give rise to numerically large contributions to the observable of interest.
Effective field theory methods are employed in order to factor the cross section into the convolution of a number of quantities
each one of which is characterized by a single physical scale. Each element in the factorization formula can be evaluated  in perturbation theory at its characteristic scale, where its perturbative expansion is free from large logarithms. Subsequently, the scale invariance of the cross section is employed in order to derive Renormalization Group Equations (RGEs) satisfied by the various elements in the factorization formula. The RGEs can then be used to run all of the elements to a common factorization scale, so that logarithmic corrections depending on large scale ratios are resummed in the evolution factors.  This approach has the advantage of trading the calculation of quantities depending on several scales with the more manageable calculation of objects depending on a single physical scale. A classic example of the procedure outlined above is represented by the calculation of the Drell-Yan cross section carried out in \cite{Becher:2007ty}; in that work, SCET methods were employed  in order to resum soft gluon emission corrections  up to next-to-next-to-next-to-leading logarithmic  accuracy.

The study of processes with colored particles in the final state is  technically more complicated because  some of the objects in the factorization formula are matrices in color space. However, due to the relevance of top-quark studies in the physics program of the Tevatron and the LHC, SCET methods were extensively employed in the study of top-quark pair production. 
 Three different singular limits of the partonic cross section for this process were considered. The first of these is the production 
threshold limit $\hat{s} \to 4 m_t^2$, where $\sqrt{\hat{s}}$ is the partonic center of mass energy and $m_t$ is the top-quark mass.
The production threshold limit is employed in order to calculate the total top-quark pair-production cross section \cite{Beneke:2009ye, Beneke:2010da, Beneke:2011mq}.
The soft  emission limit in Pair Invariant Mass kinematics (PIM), needed
for the calculation of the pair invariant-mass distribution, was considered 
 \cite{Ahrens:2009uz,Ahrens:2010zv}. 
Finally,  the soft limit in One Particle Inclusive (1PI) kinematics, which is employed  to calculate the 
 the top (or antitop) transverse momentum and rapidity distributions, was analyzed in the SCET framework in \cite{Ahrens:2011mw}. In all cases the resummation  was carried out in momentum space up to next-to-next-to-leading (NNLL) accuracy. The hard scattering kernels obtained in PIM and 1PI kinematics in \cite{Ahrens:2010zv,Ahrens:2011mw} were recently combined with semi-leptonic decays of top quarks  in a fully differential parton level Monte Carlo which allows for the study of IR safe observables constructed from momenta of top-quark decay products  \cite{Broggio:2014yca}.
Furthermore, the top-pair transverse momentum distribution at hadron colliders was resummed up to NNLL accuracy in  \cite{Zhu:2012ts, Li:2013mia}. This last observable is sensitive to small transverse momenta and its  analysis in SCET involves a collinear anomaly of the kind encountered in the study of the Drell-Yan process at small vector boson transverse momentum \cite{Becher:2010tm}.
The methods and results obtained for top-pair production can be straightforwardly  adapted to the calculation of the pair production of other colored massive particles, such as top squarks \cite{Broggio:2013cia,Broggio:2013uba},
or gluinos and squarks of the first two generations \cite{Falgari:2012sq, Beneke:2013opa}.

Here we review recent work dealing with the production of  energetic top quarks pairs, i.e. top quarks with an energy which is much larger than their mass. This kinematic region is particularly sensitive to  new physics, since many beyond the Standard Model scenarios predict the presence of new particles which decay into energetic top quarks. The characteristic signal of the existence  of these particles would be the presence of bumps or more subtle distortions in the high invariant mass region and/or high transverse momentum region of the respective differential distributions. Furthermore, boosted top quarks introduce a new level in the scale hierarchy: In this kinematic region, $\hat{s}$ is much larger than $m_t^2$. It must be observed that the results in \cite{Ahrens:2010zv,Ahrens:2011mw} were instead obtained under the assumption that $\hat{s} \sim m_t^2$. The study of the factorization of the top quark pair production cross section in PIM kinematics in the double soft emission and small top quark mass limit was initiated in \cite{Ferroglia:2012ku}. The results obtained there, complemented with the NNLO soft function for the production of massless quarks \cite{Ferroglia:2012uy}, were employed in order to derive an improved soft plus virtual NNLO approximation to the top pair invariant mass distribution \cite{Ferroglia:2013zwa}. The factorization of the top pair production cross section was studied in 1PI kinematics in \cite{Ferroglia:2013awa}. The purpose of this proceeding is to review the findings 
of \cite{Ferroglia:2012ku,Ferroglia:2012uy,Ferroglia:2013zwa,Ferroglia:2013awa}.

\section{Top pair production in the soft limit}

We start by summarizing the findings of \cite{Ahrens:2010zv,Ahrens:2011mw}. We are interested in the partonic process
\begin{equation}
p_i(p_1) + p_j(p_2) \rightarrow t (p_3) + \bar{t}(p_4) + X(k) \, ,
\end{equation}
where $p_i, p_j$ indicate the initial state partons, $X$ indicates the additional final state radiation of total momentum $k$. The partonic cross section receives  large contributions from the kinematic region in which the  final state radiation is soft. The proximity to the soft region can be parameterized by means a soft variable which vanishes in the soft limit. In the study of soft gluon emission corrections to the total pair production cross section, the soft variable is  chosen to be equal to the top quark velocity at the production threshold: $\beta = \sqrt{1 - 4 m_t^2/\hat{s}}$.
In the soft limit, $\beta \to 0$. This limit forces the produced top pair to be almost at rest. 
In order to calculate the top-pair invariant mass distribution, it is necessary to work in the framework of Pair Invariant Mass  (PIM) kinematics, where the soft limit is regulated by the parameter $z$; if one indicates the invariant mass with $M$, in the soft limit one finds
\begin{equation}
1-z \equiv 1-M^2/\hat{s} \to 0
\end{equation} 
The calculation of the top-quark transverse momentum or rapidity distribution requires the use of One Particle Inclusive (1PI) kinematics. In this case the relevant parameter is 
\begin{equation}
s_4 = (p_4 + k)^2 - m_t^2 \to 0.
\end{equation}
The soft limits in PIM and 1PI kinematics do not constrain the velocity of the produced top and antitop quarks. Furthermore, predictions for the total cross section can be obtained in PIM and 1PI kinematics by integrating the differential distributions.

When working in the soft limit, one finds a clear hierarchy among the physical scales involved in the process:
\begin{description}
\item PIM kinematics: \hspace*{0.1cm} $\hat{s}, M^2, m_t^2 \gg \hat{s} (1-z)^2 \gg \Lambda^2_{\mbox{\tiny QCD}}$\,,
\hspace*{1.1cm}  1PI kinematics: \hspace*{0.1cm}  $\hat{s}, m_t^2 \gg s_4 \gg \Lambda^2_{\mbox{\tiny QCD}}$\,.
\end{description}
It is known that in this limit the partonic cross section receives contribution exclusively from the two production channels which are already open at the tree level (quark-annihilation and gluon fusion channel). Furthermore, the partonic cross section factors into a product of a hard and a soft function; using the notation of \cite{Ahrens:2010zv,Ahrens:2011mw} one finds schematically
\begin{align}
d \hat{\sigma}_{\mbox{{\footnotesize PIM}}} &\sim \mbox{Tr}\left[ 
\bm{H}^{(m)}(M,m_t,\cos \theta,\mu) \, \bm{S}^{(m)}(\sqrt{\hat{s}} (1-z),m_t,\cos \theta,\mu) \right] +{\mathcal O}(1-z)
\nonumber \, ,\\
d \hat{\sigma}_{\mbox{{\footnotesize 1PI}}}  &\sim \mbox{Tr}\left[ 
\bm{H}^{(m)}(\hat{s},\hat{t}_1,\hat{u}_1,m_t,\mu) \, \bm{S}^{(m)}(s_4,\hat{s},\hat{t}_1,\hat{u}_1,m_t,\mu) \right]  + {\mathcal O}\left(\frac{s_4}{m_t^2}\right) , \label{eq:softfact}
\end{align}
where $\theta$ is the top quark scattering angle and the Mandelstam invariant are defined as $\hat{t}_1 = (p_1-p_3)^2-m_t^2$ and $\hat{u}_1 = (p_2-p_3)^2 -m_t^2$. 
The hard function ${\bm H} ^{(m)}$ and soft function ${\bm S}^{(m)}$ are matrices in color space. 
The superscript $(m)$ is a reminder that the functions are calculated at $m_t \neq 0$.
The hard functions receive contributions only from virtual corrections  and are identical for PIM and 1PI kinematics. The soft functions receive contributions from soft gluon emission diagrams and are different in the two kinematic schemes. 
In Eq.~(\ref{eq:softfact}) we suppressed the subscript $ij \in \{q \bar{q}, gg\}$ indicating the channel. The factorization achieves the separation of the hard and soft scales. 
If one sets the scale $\mu$ in the hard and soft functions equal to the scale that characterizes the hard process and the soft gluon emission, respectively, both functions are free from large logarithms and can be evaluated in perturbation theory.
 The calculation of the hard function and the soft function is simpler than the calculation of the full cross section. The hadronic
 cross section, obtained taking the convolution of  the partonic cross section with the PDFs, is independent from the scale $\mu$. One can exploit this fact in order to obtain Renormalization Group Equations (RGEs) satisfied by the hard functions and soft functions. It is then possible to  solve these RGEs in order to run the functions to a common factorization scale down (or up) from the scale at which they could be reliably calculated in perturbation theory. This process amounts to resumming large logarithmic corrections which depend on the ratio of hard and soft scales. 
 %
 %
 In \cite{Ahrens:2010zv,Ahrens:2011mw} all of the elements necessary in order to carry out the resummation in PIM and 1PI kinematics up to NNLL accuracy were derived.
In particular, this required the calculation of the hard and soft functions up to NLO. Alternatively, one can employ the NLO hard and soft functions in combination with their RGEs  to obtain an {\emph{approximate}} NNLO partonic cross section. The NNLO partonic cross section has the general form 
\begin{equation}
d \hat{\sigma}_{\mbox{{\footnotesize NNLO}}} =   D_3 P_3 (\lambda)  
+ D_2 P_2 (\lambda) + D_1 P_1 (\lambda)  
  + D_0 P_0 (\lambda) + C_0 \delta(\lambda) +  R(\lambda)  \, , \label{eq:approxNNLO}
\end{equation}
where $\lambda \in [1-z,s_4]$ depending on the kinematics, $P_n$ indicate the plus distributions $[\ln^n (1-z)/(1-z)]_+$
and $[\ln^n (s_4)/s_4]_+$ in PIM and 1PI kinematics, respectively, and the cofactors $D_i (i=0,\cdots,3)$ and $C_0$ are functions of the top mass and of the Mandelstam invariants. The remainders $R$  are functions which are non-singular in the soft limit. The approximate NNLO formulas derived in \cite{Ahrens:2009uz,Ahrens:2010zv, Ahrens:2011mw} include the exact analytic expression of all of the coefficients $D_i$ and the scale dependent terms in $C_0$.

Accurate predictions of the partonic cross section in the soft limit allow one to obtain reliable predictions for the hadronic cross section even if the convolution integral of PDFs and partonic cross section samples kinematics regions in which the soft emission approximation is not valid. This is due to the  steep fall-off of the PDFs away from the soft region, a phenomenon which goes under the name of  {\emph{Dynamical Threshold Enhancement}} \cite{Becher:2007ty, Bonvini:2012an}. Numerical studies reported in 
\cite{Ahrens:2009uz,Ahrens:2010zv, Ahrens:2011mw}
showed that this effect takes place in top quark pair production.
%
%
As shown in \cite{Abazov:2014vga}  and \cite{Aad:2012hg}, if one compares theoretical predictions  for the differential  distributions  at NLO+NNLL  with data one finds good agreement in both shape and normalization.



\section{Factorization for boosted top pairs in PIM kinematics}

In PIM kinematics, if one considers the situations in which the energy of the top quarks is much larger than their mass, it is necessary to study the factorization of the cross section  when, on top of the soft-emission limit hierarchy discussed before, one also assumes $\hat{s},M^2 \gg m_t^2$. 
The corresponding factorization formula, derived in \cite{Ferroglia:2012ku}, provides the framework for the simultaneous resummation of soft gluon emission corrections and of  large logarithms of the ratio $m_t/M$. The factorization of the pair invariant mass distribution for boosted top quarks was obtained by weaving together known results for the factorization in either the small mass or the soft limit in a unified description encompassing both. In particular, building upon the results of \cite{Mele:1990cw}, it was shown that, for $m_t \ll M$, the partonic cross section factors into the convolution of the cross section for \emph{massless} quark production, and a convolution of perturbative fragmentation functions for each of the heavy quarks. Given the factorized cross section in the small mass limit, it was then straightforward to add an additional layer for the soft limit in the component parts. In fact, the massless partonic cross section in the soft limit factors into  hard and soft functions, as it can be proven by means of the same methods described  in \cite{Ahrens:2010zv} for the massive case. Furthermore, the fragmentation function can be factorized into a product of collinear and soft collinear functions by using the results of \cite{Korchemsky:1992xv, Cacciari:2001cw, Cacciari:2002xb, Gardi:2005yi, Neubert:2007je}. The technical aspects of the derivation of the factorization formula are discussed in detail in \cite{Ferroglia:2012ku}; here we simply report the final result. The differential distribution in $M$ and  $\theta$ is
\begin{equation}
\frac{d^2 \sigma}{d M d \cos \theta} = \frac{8 \pi \beta_t}{s M} \sum_{ij} \int_\tau^1 \frac{dz}{z} \ff_{ij}\left(\frac{\tau}{z}, \mu_f\right) C_{ij}\left(z, M, m_t, \cos \theta, \mu_f \right) \, ,
\end{equation}
where $\sqrt{s}$ is the hadronic center of mass energy, $\tau = M^2/s$, $\beta_t = \sqrt{1 - 4 m_t^2 /M^2}$ and $\ff_{ij}$ indicates the partonic luminosity in the $ij$ channel. The Laplace transform $\tilde{c}_{ij}$ of the hard scattering kernels $C_{ij}$ in the double soft and small mass limit
factors into a product of functions as follows  
\begin{eqnarray}
\tilde{c}_{ij} \left(N, M, m_t\right) \!\!\! &=& \!\!\! C_D^2(m_t)  \mbox{Tr}\left[\! \bm{H}_{ij} \left(\!M\!\right) \bm{\tilde{s}}_{ij}\left(\!\ln\frac{M^2}{\bar{N}^2 \mu_f^2} \!\right)\!\right] \tilde{c}^{ij}_t \left(\!\ln\frac{1}{\bar{N}^2}, m_t \!\right) \tilde{s}_D^2\left(\!\ln \frac{m_t}{\bar{N} \mu_f} \!\right) + {\cal O}\left(\!\frac{1}{N},\frac{m_t}{M}\!\right) \, . \label{eq:Mfact}
\end{eqnarray}
In Eq.~(\ref{eq:Mfact}), $\bar{N} = N e^{\gamma_E}$ is the Laplace variable, $\bm{H}$ is the hard function  for the production of massless quarks,  $\bm{\tilde{s}}$ is the corresponding Laplace transformed PIM soft function, while $C_D$ and $S_D$ are the collinear and soft-collinear components of the fragmentation function, respectively. Finally, the matching coefficients $c_t^{ij}$, proportional to the number of heavy (i.e. top) flavors $n_h=1$, account for the fact that partonic luminosities and fragmentation functions are calculated by considering $n_l =5$ active flavors, while $\bm{H}$ and $\bm{\tilde{s}}$ are calculated in a theory with $n_l+n_h$ active flavors. The arguments $\cos \theta$ in $\tilde{c}$, $\hat{t}_1$ in $\bm{H}$ and $\bm{\tilde{s}}$, and $\mu$ in all functions have been suppressed in Eq.~(\ref{eq:Mfact}).
A resummed cross section appropriate for the both the soft emission and small mass limit can then be obtained by solving the RGEs for the different component functions separately. The anomalous dimensions appearing in the RGEs are known to an order which is  sufficient to implement the resummation of both soft and mass logarithms to NNLL accuracy.

While in order to carry out NNLL resummation one needs to know the factors in Eq.~(\ref{eq:Mfact}) to NLO only, all of these elements are known up to NNLO. In particular, the perturbative fragmentation function was calculated up to NNLO in \cite{Melnikov:2004bm} and the NNLO soft functions were evaluated in \cite{Ferroglia:2012uy}. 
Very recently, the NNLO hard functions for all of the $2 \to 2$ processes in massless QCD were evaluated in \cite{Broggio:2014xxx},
although the information needed in what follows was at first obtained in a different way  (see \cite{Ferroglia:2013zwa} and references therein).
 With these elements it is then possible to assemble a complete soft plus virtual approximation to the NNLO cross section
 \cite{Ferroglia:2013zwa}, valid in the double soft emission and small mass limit. This means that one can obtain the 
 coefficients $D_i$ and $C_0$ in Eq.~(\ref{eq:approxNNLO}) up to very small terms suppressed by positive powers of $m_t/M$.
 This does not add anything to the knowledge of the coefficients $D_i$, whose exact $m_t$ dependence was already derived in
 \cite{Ahrens:2009uz, Ahrens:2010zv} starting from the NNLL resummation formula in the soft limit only. Nevertheless, the agreement between the results of \cite{Ferroglia:2013zwa}
 and the $m_t \to 0$ limit of the coefficients $D_i$ found in \cite{Ahrens:2009uz, Ahrens:2010zv} represents a stringent test on the factorization scheme of \cite{Ferroglia:2012ku}.
 Furthermore, one can obtain 
 useful information on the coefficient $C_0$, which has the following schematic structure 
\begin{equation}
C_0 = C_{0,2} \ln^2 \frac{m_t^2}{M^2} +  C_{0,1} \ln \frac{m_t^2}{M^2} + C_{0,0} + {\cal O}\left(\frac{m_t^2}{M^2}\right) \, ;
\end{equation}
indeed, the coefficients $C_{0,i}$ were completely determined in \cite{Ferroglia:2013zwa}. By combining the $D_i$ exact in $m_t$ and the $m_t \to 0$ limit of $C_0$, one obtains an improved approximate NNLO formula for the pair invariant mass distribution which is more complete than the approximate NNLO result  obtained  in \cite{Ahrens:2009uz, Ahrens:2010zv}. 

The phenomenological impact of these improved approximate NNLO corrections was studied in \cite{Ferroglia:2013zwa}. It was found that the new NNLO approximation produces moderate enhancements of the differential cross section when compared with the results of \cite{Ahrens:2009uz, Ahrens:2010zv}. It is interesting to compare the total cross section obtained by integrating the approximate NNLO formulas of \cite{Ahrens:2009uz, Ahrens:2010zv}, (which we will refer to as NNLO approx. PIM), the cross section obtained by integrating the approximate NNLO
formulas of  \cite{Ferroglia:2013zwa} (indicated by NNLO approx. C), and the exact NNLO corrections to the total cross section, which were evaluated in \cite{Czakon:2013goa}. Predictions based upon NNLO approx. PIM are smaller than exact NNLO calculations both at the Tevatron and at the LHC. Predictions based upon NNLO approx. C are still smaller than the exact result but the additional terms included in the delta function coefficient help to bridge the gap between the exact NNLO result and NNLO approx. PIM calculations. In particular, the range of values determined by varying the factorization/renormalization scale in NNLO approx. C has a sizable overlap with the corresponding uncertainty range in the complete NNLO result. These conclusions can be drawn by looking at Table~5, in \cite{Ferroglia:2013zwa}.

In \cite{Ferroglia:2013zwa} the NNLO approx. C results for the pair invariant mass distribution cross section were compared with results including NNLL resummation \cite{Ahrens:2010zv}. It was possible to conclude that at relatively low values of the invariant mass, where the cross section is large, resummation is only a small effect and fixed-order perturbation theory is sufficient. However, for larger values of the invariant mass NNLL corrections are quite large and provide a further enhancement of the cross section; for example, at LHC with a center of mass energy of $7$ TeV, at $M = 3$ TeV NNLO corrections are roughly of the same size as NLO ones, while NNLL resummation corrections are even larger. In that kinematic region fixed-order perturbation theory breaks down, and the resummation of soft gluon emission effects is necessary.

\section{Factorization for boosted top pairs in 1PI kinematics}

The study of the double soft emission and small mass limit in 1PI kinematics was carried out in \cite{Ferroglia:2013awa}.  The 
relevant scale hierarchy is assumed to be $\hat{s} \gg s_4 \gg \Lambda^2_{\mbox{\tiny QCD}}$ and $\hat{s} \gg m_t^2$. In this case, it is convenient to start the study of the double  limit from the factorized form of the partonic cross section in the soft emission limit, Eq.~(\ref{eq:softfact}). One can then consider the $m_t \to 0$ limit of the function $\bm{H}^{(m)}$ and $\bm{S}^{(m)}$. In the small mass limit the $\bm{H}^{(m)}$ factors into the product of the hard function for the production of massless top-quarks, $\bm{H}$, and a function encoding all of the $m_t$ dependence which is related to collinear divergences in the small mass limit, $C_D$. Both of  these elements are 
already found in the PIM factorization formula, Eq.~(\ref{eq:Mfact}). On the other hand, the small mass factorization of $\bm{S}^{(m)}$ is very different from the PIM case. The massive soft function factors into the product of three component functions, related to soft radiation collinear to the top quark, soft radiation collinear to the unobserved antitop quark, and wide-angle soft emission, respectively. In \cite{Ferroglia:2013awa}, the three functions were defined at the operatorial level in terms of Wilson loops. The wide-angle soft emission is associated with a Wilson loop built out of four light-like Wilson lines and involving a delta function constraint dictated by 1PI kinematics. Soft radiation collinear to the top quark is associated with a Wilson loop defining the soft part of the heavy quark fragmentation function, which is the object defined in \cite{Mele:1990cw}. Finally, soft radiation collinear to the antitop quark is associated to the Wilson loop defining the heavy-quark jet function introduced in \cite{Fleming:2007qr}. 
%
%
The double differential cross section in 1PI kinematics reads
\begin{equation}
\frac{d \sigma}{d p_T dy} = \frac{16 \pi p_T}{3 s} \sum_{ij} \int_{x_1^{\mbox{{\tiny min}}}}^1 \frac{d x_1}{x_1} \int_{x_2^{\mbox{{\tiny min}}}}^1 \frac{d x_2}{x_2} f_{i/N_1} (x_1, \mu_f) f_{j/N_2} (x_2, \mu_f) C_{ij} \left(s_4,\hat{s},\hat{t}_1, \hat{u}_1, m_t, \mu_f \right) \, , 
\end{equation}
where $p_T$ and $y$ represent the top-quark transverse momentum and rapidity while the extrema of the integration over the energy fractions $x_1, x_2$ can be found in \cite{Ferroglia:2013awa}. The factored form of the Laplace transformed 1PI hard scattering coefficients $C_{ij}$ is 
\begin{equation}
\label{eq:Laplfac}
\tilde{c}_{ij}(N,\hat{s},m_t) \!  
= \!   C_D^2\left(\!\ln\frac{m_t^2}{\mu^2}\!\right) \!\mbox{Tr} 
\left[\!\bm{H}_{ij}\left(\!\ln\frac{\hat{s}}{\mu^2}\!\right) \, \bm{\tilde{s}}_{ij}\left(\!\ln\frac{\hat{s}}{\bar{N}^2 \mu^2} \!\right) \!\right] \!  \tilde{s}_D \left(\!\ln\frac{m_t}{\bar{N} \mu} \!\right) 
\!\tilde{s}_B \left(\!\ln \frac{\hat{s}}{\bar{N} m_t \mu} \!\right) + \mathcal{O}\left(\!\frac{\hat{s}}{N m_t^2},\frac{m_t^2}{\hat{s}}\!\right) \,.
\end{equation}
 The arguments $\hat{t}_1, \hat{u}_1$ (which should be included in the function $\tilde{c}$), $\hat{t}_1/\hat{s}$ (in the functions $\bm{H}$ and  $\bm{\tilde{s}}$), and  $\mu$ (which should be present in all functions) are not explicitly written in Eq.~(\ref{eq:Laplfac}). The color space matrix $\bm{\tilde{s}}$ describes wide-angle emission, $\tilde{s}_D$ is connected 
to emission collinear to the top-quark and it is the same as in  the PIM factorization formula, while $\tilde{s}_B$ is connected to emission collinear to the antitop quark. NNLO virtual corrections involving a closed heavy-quark loop are neglected in Eq.~(\ref{eq:Laplfac}).

As for the PIM case, the factorization formula can be employed either to resum soft and small-mass logarithms in the differential partonic cross section by solving the RGEs for the five component functions mentioned above, or  to derive improved approximate NNLO formulas for the top-quark transverse momentum or rapidity distributions.
Out of the five component functions in Eq.~(\ref{eq:Laplfac}), four were calculated to NNLO before the work on \cite{Ferroglia:2013awa}; besides for the functions already present in the PIM factorization formula, $\tilde{s}_B$ is easily derived from the results in \cite{Jain:2008gb} up to NNLO. The 1PI soft function for the production of massless quarks was calculated to NNLO in \cite{Ferroglia:2013awa}.
As for the case of PIM kinematics, with these ingredients at hand it was possible to assemble an almost complete soft plus virtual approximation to the NNLO $p_T$ and rapidity distribution \cite{Ferroglia:2013awa}, which goes beyond the approximate NNLO calculations based solely on soft gluon emission factorization carried out in \cite{Ahrens:2011mw}. A full phenomenological analysis of the improved approximate NNLO formulas for the pair invariant mass and transverse momentum distributions, as well as a complete numerical analysis of the simultaneous resummation of soft emission and small mass logarithms up to  NNLL accuracy,  was postponed to future work.

\Acknowledgements
The work of A.F. was supported in part by the PSC-CUNY award 67616-00 45 and by the NSF Award PHY-1068317.


\begin{thebibliography}{99}


\bibitem{Becher:2007ty} 
  T.~Becher, M.~Neubert and G.~Xu,
  JHEP {\bf 0807}, 030 (2008)

\bibitem{Beneke:2009ye} 
  M.~Beneke, M.~Czakon, P.~Falgari, A.~Mitov and C.~Schwinn,
  Phys.\ Lett.\ B {\bf 690}, 483 (2010)

\bibitem{Beneke:2010da} 
  M.~Beneke, P.~Falgari and C.~Schwinn,
  Nucl.\ Phys.\ B {\bf 842}, 414 (2011)
  
  \bibitem{Beneke:2011mq} 
    M.~Beneke, P.~Falgari, S.~Klein and C.~Schwinn,
    Nucl.\ Phys.\ B {\bf 855}, 695 (2012)
    
\bibitem{Ahrens:2009uz} 
  V.~Ahrens, A.~Ferroglia, M.~Neubert, B.~D.~Pecjak and L.~L.~Yang,
  Phys.\ Lett.\ B {\bf 687}, 331 (2010)
  
\bibitem{Ahrens:2010zv} 
  V.~Ahrens, A.~Ferroglia, M.~Neubert, B.~D.~Pecjak and L.~L.~Yang,
  JHEP {\bf 1009}, 097 (2010)
  
  
  \bibitem{Ahrens:2011mw} 
    V.~Ahrens, A.~Ferroglia, M.~Neubert, B.~D.~Pecjak and L.~L.~Yang,
    JHEP {\bf 1109}, 070 (2011)

\bibitem{Broggio:2014yca} 
  A.~Broggio, A.~S.~Papanastasiou and A.~Signer,
  arXiv:1407.2532 [hep-ph].
  
  \bibitem{Zhu:2012ts} 
    H.~X.~Zhu, C.~S.~Li, H.~T.~Li, D.~Y.~Shao and L.~L.~Yang,
    Phys.\ Rev.\ Lett.\  {\bf 110}, 082001 (2013)
    
\bibitem{Li:2013mia} 
  H.~T.~Li, C.~S.~Li, D.~Y.~Shao, L.~L.~Yang and H.~X.~Zhu,
  Phys.\ Rev.\ D {\bf 88}, 074004 (2013)
  
  \bibitem{Becher:2010tm} 
    T.~Becher and M.~Neubert,
    Eur.\ Phys.\ J.\ C {\bf 71}, 1665 (2011)

\bibitem{Broggio:2013cia} 
  A.~Broggio, A.~Ferroglia, M.~Neubert, L.~Vernazza and L.~L.~Yang,
  JHEP {\bf 1403}, 066 (2014)

\bibitem{Broggio:2013uba} 
  A.~Broggio, A.~Ferroglia, M.~Neubert, L.~Vernazza and L.~L.~Yang,
  JHEP {\bf 1307}, 042 (2013)

\bibitem{Falgari:2012sq} 
  P.~Falgari, C.~Schwinn and C.~Wever,
  JHEP {\bf 1301}, 085 (2013)

\bibitem{Beneke:2013opa} 
  M.~Beneke, P.~Falgari, J.~Piclum, C.~Schwinn and C.~Wever,
  PoS RADCOR {\bf 2013}, 051 (2013)
  
  \bibitem{Ferroglia:2012ku} 
    A.~Ferroglia, B.~D.~Pecjak and L.~L.~Yang,
    Phys.\ Rev.\ D {\bf 86}, 034010 (2012)


\bibitem{Ferroglia:2012uy} 
  A.~Ferroglia, B.~D.~Pecjak, L.~L.~Yang, B.~D.~Pecjak and L.~L.~Yang,
  JHEP {\bf 1210}, 180 (2012)

\bibitem{Ferroglia:2013zwa} 
  A.~Ferroglia, B.~D.~Pecjak and L.~L.~Yang,
  JHEP {\bf 1309}, 032 (2013)

\bibitem{Ferroglia:2013awa} 
  A.~Ferroglia, S.~Marzani, B.~D.~Pecjak and L.~L.~Yang,
  JHEP {\bf 1401}, 028 (2014)
 
  \bibitem{Bonvini:2012an} 
    M.~Bonvini, S.~Forte and G.~Ridolfi,
    Phys.\ Rev.\ Lett.\  {\bf 109}, 102002 (2012)
    
\bibitem{Abazov:2014vga} 
  V.~M.~Abazov {\it et al.}  [D0 Collaboration],
  arXiv:1401.5785 [hep-ex].

\bibitem{Aad:2012hg} 
  G.~Aad {\it et al.}  [ATLAS Collaboration],
  Eur.\ Phys.\ J.\ C {\bf 73}, 2261 (2013)

\bibitem{Mele:1990cw} 
  B.~Mele and P.~Nason,
  Nucl.\ Phys.\ B {\bf 361}, 626 (1991).

\bibitem{Korchemsky:1992xv} 
  G.~P.~Korchemsky and G.~Marchesini,
  Nucl.\ Phys.\ B {\bf 406}, 225 (1993)
  
  \bibitem{Cacciari:2001cw} 
    M.~Cacciari and S.~Catani,
    Nucl.\ Phys.\ B {\bf 617}, 253 (2001)

\bibitem{Cacciari:2002xb} 
  M.~Cacciari and E.~Gardi,
  Nucl.\ Phys.\ B {\bf 664}, 299 (2003)

\bibitem{Gardi:2005yi} 
  E.~Gardi,
  JHEP {\bf 0502}, 053 (2005)

\bibitem{Neubert:2007je} 
  M.~Neubert,
  arXiv:0706.2136 [hep-ph].

\bibitem{Melnikov:2004bm} 
  K.~Melnikov and A.~Mitov,
  Phys.\ Rev.\ D {\bf 70}, 034027 (2004)


\bibitem{Broggio:2014xxx}
A.~Broggio, A.~ Ferroglia, B.~Pecjak, and Z.~Zhang, in preparation.

\bibitem{Czakon:2013goa} 
  M.~Czakon, P.~Fiedler and A.~Mitov,
  Phys.\ Rev.\ Lett.\  {\bf 110}, no. 25, 252004 (2013)

  \bibitem{Fleming:2007qr} 
    S.~Fleming, A.~H.~Hoang, S.~Mantry and I.~W.~Stewart,
    Phys.\ Rev.\ D {\bf 77}, 074010 (2008)

\bibitem{Jain:2008gb} 
  A.~Jain, I.~Scimemi and I.~W.~Stewart,
  Phys.\ Rev.\ D {\bf 77}, 094008 (2008)



\end{thebibliography}
\end{document}